\documentclass[aps,fleqn,superscriptaddress]{revtex4}
\pdfoutput=1
\usepackage{graphicx,color,natbib}
\usepackage{amsmath,amssymb,amsfonts}

\newcommand{\bse}{\begin{subequations}}
\newcommand{\ese}{\end{subequations}}
\newcommand{\be}{\begin{equation}}
\newcommand{\ee}{\end{equation}}
\newcommand{\bea}{\begin{eqnarray}}
\newcommand{\eea}{\end{eqnarray}}
\newcommand{\ba}{\begin{array}}
\newcommand{\ea}{\end{array}}

\def\m{\frac{\mu}{T}}
\input amssym.def
\input amssym.tex

\usepackage[colorlinks=true, linkcolor=blue, bookmarks=true]{hyperref}

\begin{document}

\title{Holographic Entanglement of Purification near a Critical Point}

\author{B. Amrahi\footnote{ b$_{-}$amrahi@sbu.ac.ir}}
\affiliation{Department of Physics, Shahid Beheshti University G.C., Evin, Tehran 19839, Iran}
\author{M. Ali-Akbari\footnote{$\rm{m}_{-}$aliakbari@sbu.ac.ir}}
\affiliation{Department of Physics, Shahid Beheshti University G.C., Evin, Tehran 19839, Iran}
\author{M. Asadi\footnote{$\rm{m}_{-}$asadi@ipm.ir}}
\affiliation{School of Physics, Institute for Research in Fundamental Sciences (IPM),
P.O.Box 19395-5531, Tehran, Iran}

\begin{abstract}
In the presence of finite chemical potential $\mu$, we holographically compute the entanglement of purification in a $2+1$- and $3+1$-dimensional field theory and also in a $3+1$-dimensional field theory with a critical point. We observe that compared to $2+1$- and $3+1$-dimensional field theories,  the behavior of entanglement of purification near critical point is different and it is not a monotonic function of $\m$ where $T$ is the temperature of the field theory. Therefore, the entanglement of purification distinguishes the critical point in the field theory. We also discuss the dependence of the holographic entanglement of purification on the various parameters of the theories. Moreover, the critical exponent is calculated.
\end{abstract}

\maketitle

\tableofcontents

\section{Introduction}
The AdS/CFT correspondence or more generally gauge-gravity duality (as an example of the holographic idea) opens a new window to study strongly coupled field theories. This duality enables us to describe and study various phenomena in the field theories utilizing their corresponding gravity duals. Although these phenomena may not seem simple from a field theory point of view, their gravitational descriptions hopefully have a simpler explanation. As an example, the confinement-deconfinement phase transition of quantum chromodynamics at low energy was highly discussed in the literature and it corresponds to the Hawking-Page phase transition on the gravity side \cite{CasalderreySolana:2011us}. Therefore, based on this idea the gravitational counterparts for different quantities in the field theory have been defined and thereby various properties of the field theory have been investigated. 
As another example, in the context of quantum information theory, entanglement entropy is one of the most well-known quantities which has simple holographic dual. Entanglement entropy determines quantum entanglement between subsystems $A$ and its complementary for a given {\it{pure}} state. On the gravity side, it corresponds to the area of the minimal surface with a suitable condition at the boundary which is usually called RT-surface \cite{Ryu:2006bv, Ryu:2006ef, Hubeny:2007xt}. This prescription has been frequently discussed and successfully passed a lot of non-trivial tests. Indeed, since in order to calculate entanglement entropy one only needs to compute an area, the calculation is much simpler in the gravity theory than in the strongly coupled field theory. 
A new quantity which recently received a lot of interest in the gauge-gravity duality point of view is entanglement of purification (EoP) $E_p$. It measures correlations (quantum and classical) between two disjoint subsystems $A$ and $B$ for a given {\it{mixed}} state described by density matrix $\rho_{AB}$ where $AB=A \cup B$. Then, the EoP is defined as
\be\label{eof}
E_p(\rho_{AB})=\underset{\rho_{AB}=Tr_{A'B'}|\psi\rangle\langle\psi|}{\rm{min}} S_{AA'},
\ee
where $\rho_{AA'}=Tr_{BB'}|\psi\rangle\langle\psi|$, $S_{AA'}$ is the entanglement entropy associated with $\rho_{AA'}$ and the state $|\psi\rangle$ satisfies in the following conditions
\begin{itemize}
\item $|\psi\rangle \in {\cal{H}}_{AA'}\otimes{\cal{H}}_{BB'}$ where $A'$ and $B'$ are arbitrary. In fact by adding auxiliary degrees of freedom we construct a pure state, i.e. $|\psi\rangle$.
\item $|\psi\rangle$s are pure states satisfy the condition $\rho_{AB}=Tr_{A'B'}|\psi\rangle\langle\psi|$ and therefore they are called purification of $\rho_{AB}$. 
\end{itemize}
Note that the minimization in \eqref{eof} is taken over any pure states $|\psi\rangle$. It is then conjectured that the EoP is holographically dual to entanglement wedge cross-section $E_w$ of $\rho_{AB}$, as a measure of correlation between $A$ and $B$, which is defined \cite{Takayanagi:2017knl, Nguyen:2017yqw} 
\be\label{eop} %
E_w(\rho_{AB})=\frac{{\rm{Area}}(\Sigma_{AB}^{min})}{4G_N},
\ee 
where $\Sigma_{AB}^{min}$ is the minimal area surface in the entanglement wedge $E_w(\rho_{AB})$ that ends on the RT-surface of $A\cup B$, the green line in figure \ref{fig2} and $G_N$ is Newton's constant.
As a result, we have 
\be 
E_p(\rho_{AB})=E_w(\rho_{AB}).
\ee 
Using this prescription, all properties of EoP can be described holographically, for instance, see \cite{Takayanagi:2017knl}. Moreover, it is also discussed that the EoP experiences a discontinuous transition when the two subsystems under study are distant enough. In addition, the above proposal generalizes to the time-dependent case and thereby quantum quenches have been studied \cite{Yang:2018gfq}. Furthermore, in order to understand various aspects of the EoP, many papers appear in the literature, for example, see \cite{Bhattacharyya:2018sbw, Espindola:2018ozt, Umemoto:2018jpc, Liu:2019qje, BabaeiVelni:2019pkw}.
\section{Calculating of the EoP in our models }
In this paper, to compute the EoP, we would like to start with a general metric 
\be \label{generalmetric}
ds^2=f_1(r)dt^2+f_2(r)dr^2+f_3(r)d\vec{x}^2,
\ee %
where $\vec{x}\equiv (x_1,...,x_d)$. The above metric is asymptotically AdS$_{d+2}$ and $r$ is the radial direction. The strongly coupled field theory lives on the boundary at $r\rightarrow \infty$. $f_1$, $f_2$ and $f_3$ are arbitrary functions and will be fixed later on. We then consider two subsystems $A$ and $B$ at a given time slice as follows
\be\begin{split} %
x(r) &\equiv x_1(r), \ {\rm{where}}\ x(r)\ {\rm{is\ an\ odd\ function\ of\ }} r,\cr
-\frac{L}{2}&\leq x_i \leq \frac{L}{2},\ \ \ i=2,...,d,
\end{split}\ee %
\begin{figure}
\centering
\includegraphics[width=100 mm]{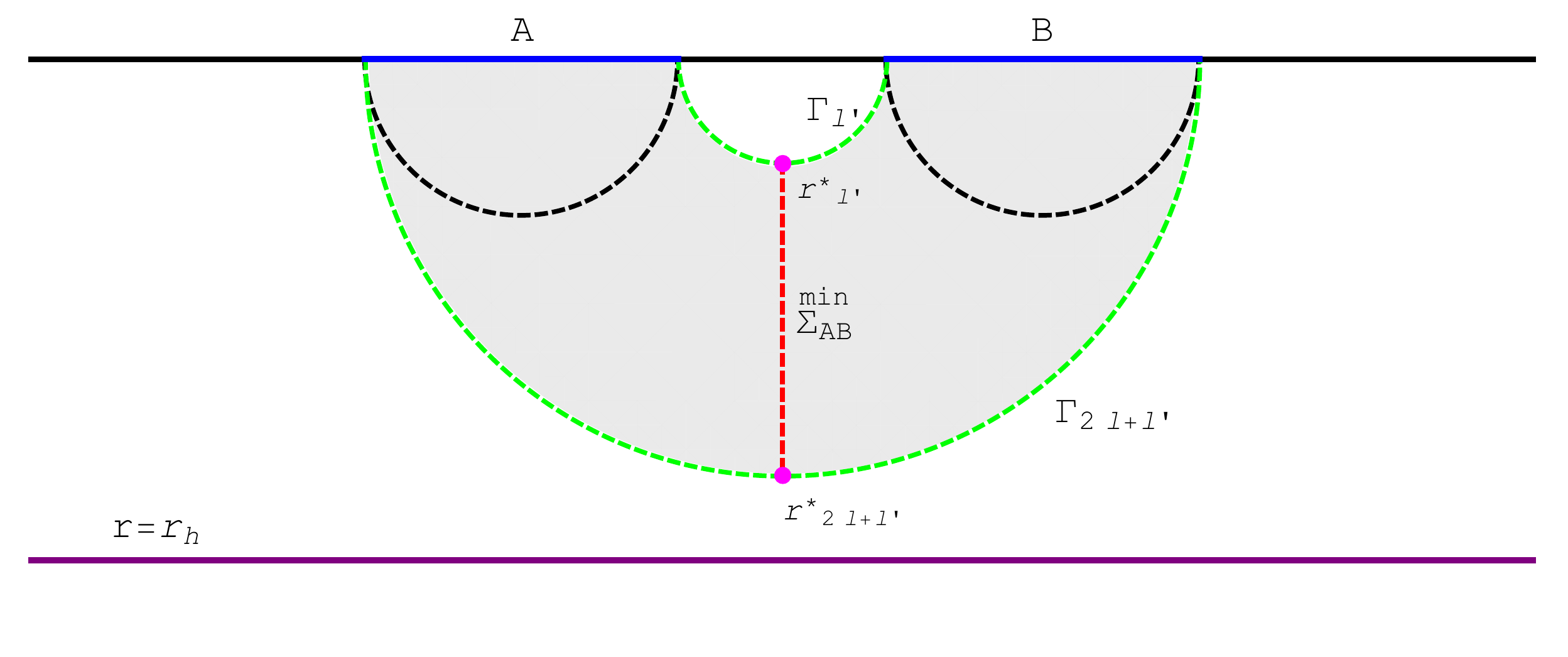} 
\caption{The gray region shows the entanglement wedge dual to $\rho_{AB}$. The minimal surfaces, RT-surfaces, are denoted by $\Gamma$, the dashed curves.}
\label{fig2}
\end{figure} 
where the length of the both disjoint subsystems is equal to $l$ and $l'$ is distance between two subsystems, see figure \ref{fig2}. Thus for the case at hand it is easy to see that $\Sigma_{AB}^{min}$ runs along the radial direction and connects the minimum point of minimal surfaces $\Gamma_{l'}$ and $\Gamma_{2l+l'}$. Then, using \eqref{eop}, the area of this hypersurface turns out to be 
\be\label{heop}
E_w=\frac{L^{d-1}}{4G_N}\int_{r^*_{2l+l'}}^{r^*_{l'}}dr \sqrt{f_2 f_3^{d-1}},
\ee
where $r^*_{l'}$ and $r^*_{2l+l'}$ denote the turning point of $\Gamma_{l'}$ and $\Gamma_{2l+l'}$, respectively. Clearly, due to the even symmetry of $x(r)$, the final result for the EoP does not depend on $x$. In order to find the value of the turning point, let's say $r^*_{l'}$, we calculate the area of the following configuration 
\be\begin{split} %
-\frac{l'}{2}&\leq x(r) \equiv x_1(r)\leq \frac{l'}{2},\cr
-\frac{L}{2}&\leq x_i \leq \frac{L}{2},\ \ \ i=2,...,d,
\end{split}\ee %
which leads to 
\be 
{\rm{Area}}=2L^{d-1}\int_{r^*_{l'}}^\infty  dr f_3^{\frac{d-1}{2}} \sqrt{f_2-f_3\ x'(r)^2}\equiv \int dr {\cal{L}}.
\ee 
Since ${\cal{L}}$ does not depend on $x$ explicitly, the corresponding Hamiltonian is constant and it is then easy to find 
\be %
\frac{l'}{2}=\int_{r^*_{l'}}^\infty dr \sqrt{\frac{f_2f_{3*}^d}{f_3(f_{3}^d-f_{3*}^d)}},
\ee %
where the constant is chosen to be $\sqrt{\frac{f_2(f_3^d-f_{3*}^d)}{f_3}}$. The above equation, for a given value of $l'$, can be used to find $r^*_{l'}$, at least numerically. Similarly, one can find the value of $r^*_{2l+l'}$.
Now we are interested in studying the EoP near a critical point. Therefore, the metric we consider here, in the AdS radius unit, is 
\be \label{metric}
f_1(r)=-e^{2A(r)}h(r),\ \ f_2(r)=\frac{e^{2B(r)}}{h(r)},\ \ f_3(r)=e^{2A(r)},
\ee 
where in this case $d=3$ and 
\be \begin{split}
A(r)=\ln \left(r(1+\frac{Q^2}{r^2})^{\frac{1}{6}}\right),\ \ B(r)=-\ln \left(r(1+\frac{Q^2}{r^2})^{\frac{1}{3}}\right),\ \ 
h(r)=1-\frac{M^2}{r^2(r^2+Q^2)}.
\end{split}\ee %
This metric describes a charged black hole background. $M$ is the black hole mass and $Q$ is its charge and its horizon is located at $r_h$. The latter can be obtained from $h(r_h)=0$, leading to 
\be
r_h=\sqrt{\frac{\sqrt{Q^4+4M^2}-Q^2}{2}}.
\ee
The temperature $T$ and chemical potential $\mu$ of the field theory, which is dual to metric \eqref{metric}, are given by 
\be\begin{split} 
T&=\frac{2r_h^2+Q^2}{2\pi\sqrt{Q^2+r_h^2}},\cr 
\mu&=\frac{Q r_h}{\sqrt{Q^2+r_h^2}}.
\end{split}\ee
Since the underling theory is conformal, all physical quantities can be expressed as a function of dimensionless parameter $\m$. It was then shown that there is a critical point at $\m=(\m)_*=\frac{\pi}{\sqrt{2}}\ (\frac{Q}{r_h}=\sqrt{2})$ and the solutions are thermodynamically stable for $\frac{Q}{r_h}<\sqrt{2}$.
Using \eqref{heop} and \eqref{metric}, one can easily find 
\be\label{EoP1}
E_p\equiv \frac{4G_N}{L^{2}}E_w=\int_{r^*_{2l+l'}}^{r^*_{l'}} dr \frac{r}{\sqrt{1-\frac{M^2}{r^2(r^2+Q^2)}}}.
\ee

In order to study the behavior of the EoP near the critical point and compare these results to the results for the field theory without critical point,
we now consider RN-AdS$_{d+2}$ metric which can be written as \cite{Galante:2012pv}
\be\begin{split}\label{metric1}
ds^2&=-r^2 f(r) dt^2+\frac{1}{r^2f(r)}dr^2+r^2 d\vec{x}^2, \cr
f(r)&=1 - \frac{M}{r^{d+1}} + \frac{Q^2}{r^{2d}},
\end{split}\ee
in the AdS radius unit. Comparing to \eqref{generalmetric}, we have 
\be %
f_1(r) = r^2 f(r)=\frac{1}{f_2(r)},\ \ f_3(r)=r^2.
\ee %
Moreover, the above background contains a time component of the gauge field introduced. $M$ and $Q$ are the mass and the charge of the RN-AdS black hole, respectively. As usual, $r$ is radial coordinate and $r\rightarrow \infty$ is the AdS boundary. In addition, $\vec{x}$ are the $d$ dimensional coordinates at the boundary. The gauge-gravity duality indicates that the Hawking temperature of the black hole corresponds to the temperature of the gauge theory. The temperature of the RN-AdS$_{d+2}$ black hole is
\be\label{tem}
T = \frac{r_h}{4\pi} \bigg((d+1) -(d-1) \frac{ Q^2 }{r_h^{2d}}\bigg).
\ee%
Here $r_h$ is the radius of the event horizon, i.e. the largest root of $f(r)=0$. The relation between $r_h$, $M $ and $Q$ is
\be\label{massradius}
M = r_h^{d+1}+\frac{Q^2}{r_h^{d-1}}~.
\ee
Moreover, the gauge-gravity duality provides a correspondence between the time component of gauge field at the boundary and the chemical potential in dual boundary gauge theory. Therefore, it turns out to be \cite{Galante:2012pv, Myers:2009ij} i.e.
\be\label{chemical100}
{\mu} =\sqrt{\frac{d}{2(d-1)}} \frac{Q}{r_h^{d-1}}.
\ee
Hence, it is easy to find that 
\be
\frac{\mu}{T} = \frac{1}{\sqrt{2(d-1)}}\frac{4\pi\sqrt{d}Qr_h^d}{(d+1)r_h^{2d}-(d-1)Q^2}.
\ee
Using \eqref{heop} and \eqref{metric1}, one can easily find 
\be\label{EoP2}
E_p\equiv \frac{4G_N}{L^{2}}E_w=\int_{r^*_{2l+l'}}^{r^*_{l'}} dr \frac{r^{d-2}}{\sqrt{1-\frac{M}{r^{d+1}}+\frac{Q^2}{r^{2d}}}}.
\ee
Evidently, \eqref{EoP1} and \eqref{EoP2} reduce to the same equation in the limit of $Q\rightarrow 0$ and $d=3$.
\begin{figure}
\centering
\includegraphics[width=60 mm]{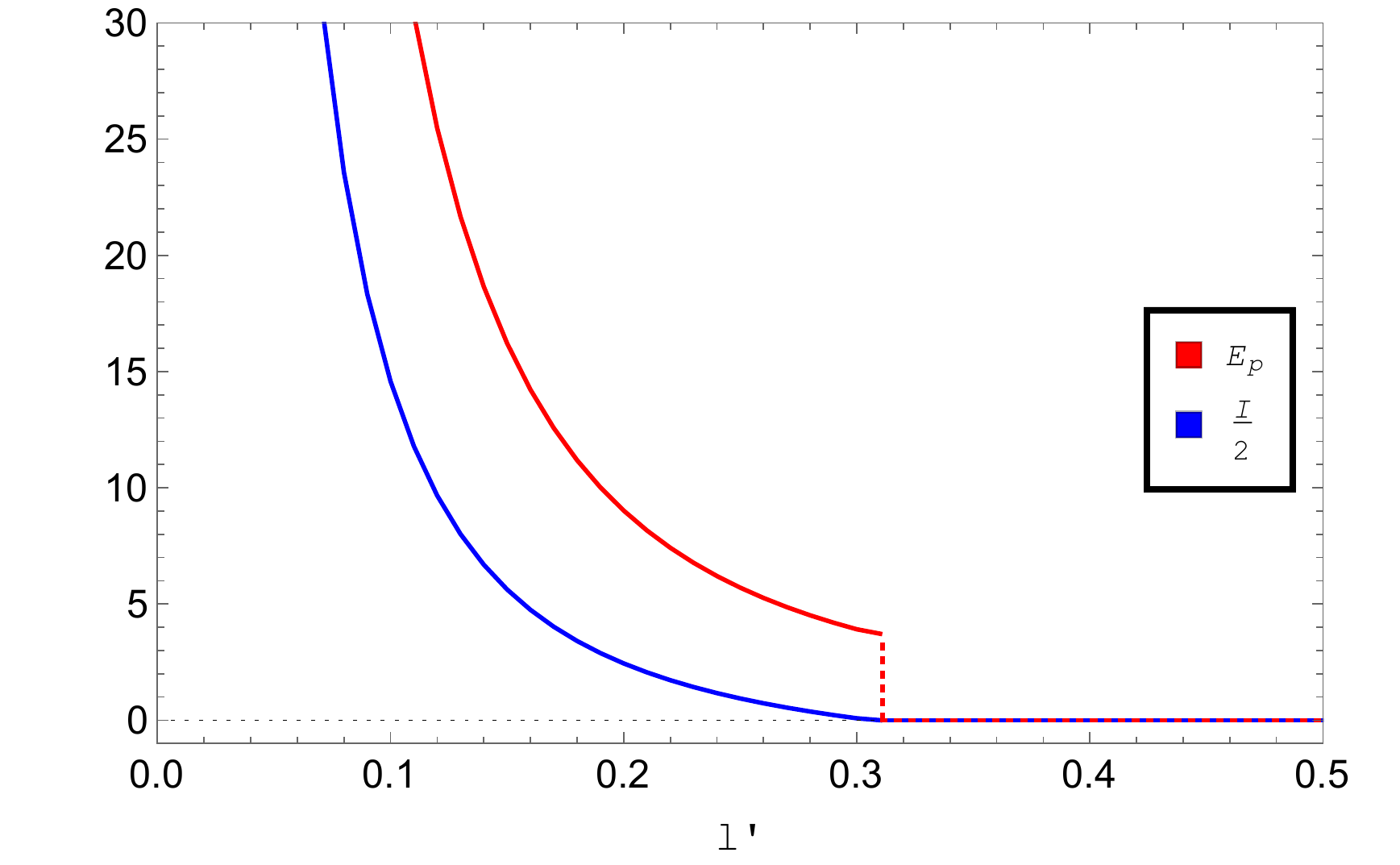}
\includegraphics[width=60 mm]{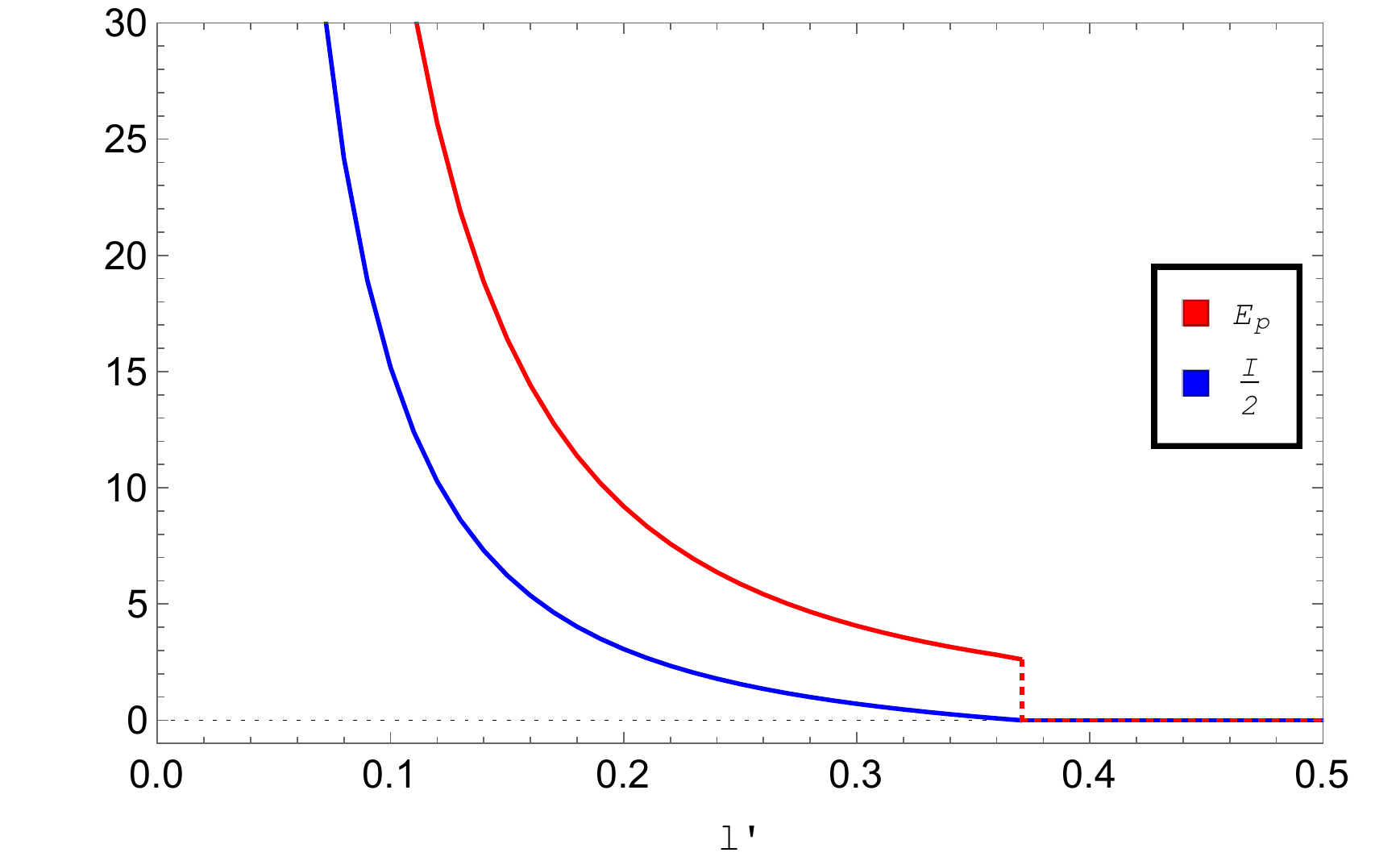} 
\caption{The EoP and $I/2$ with respect to $l'$ for $l=0.5$ (left) and $l=0.8$ (right).}
\label{fig4}
\end{figure} 
\begin{figure}
\centering
\includegraphics[width=60 mm]{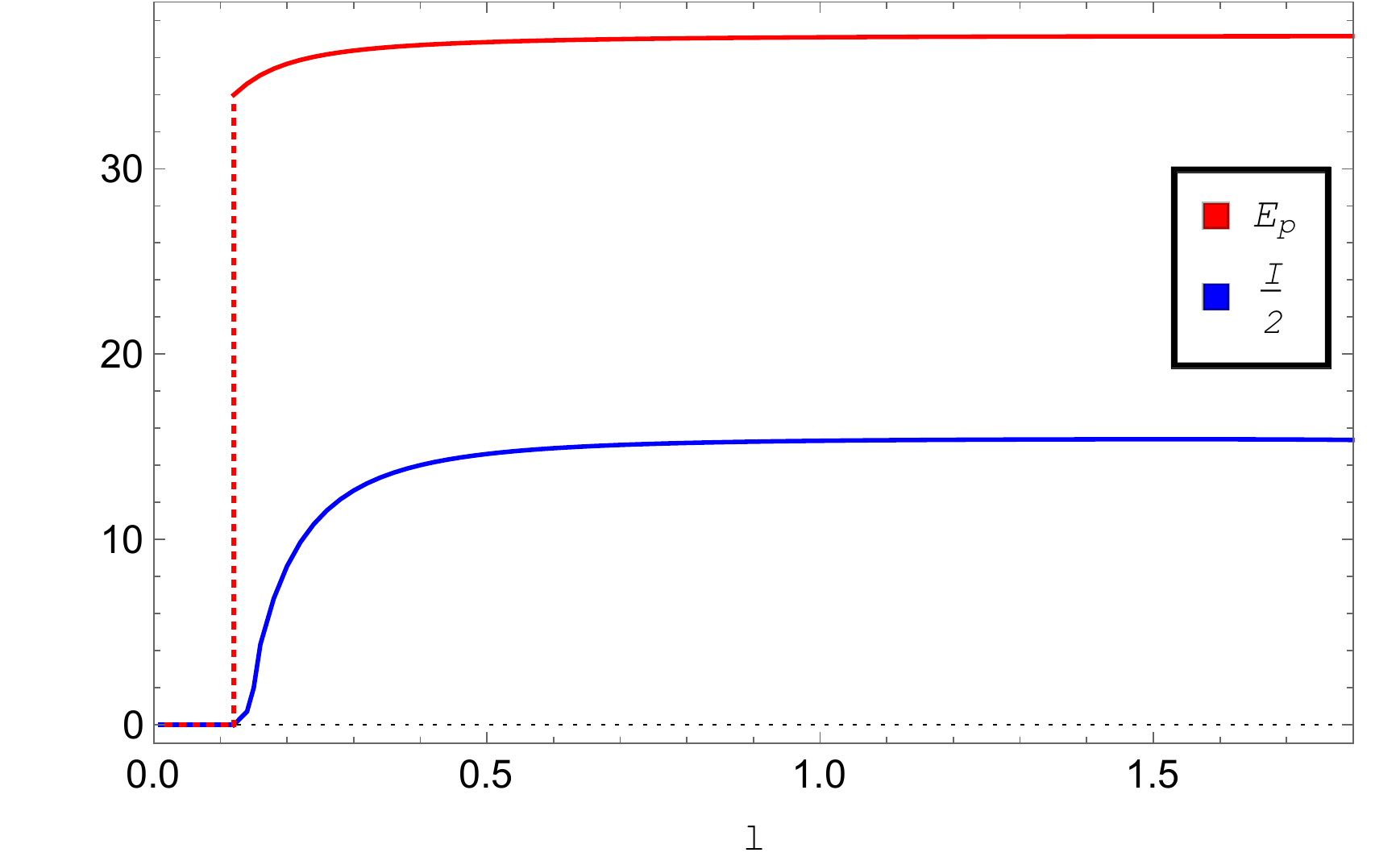} 
\includegraphics[width=60 mm]{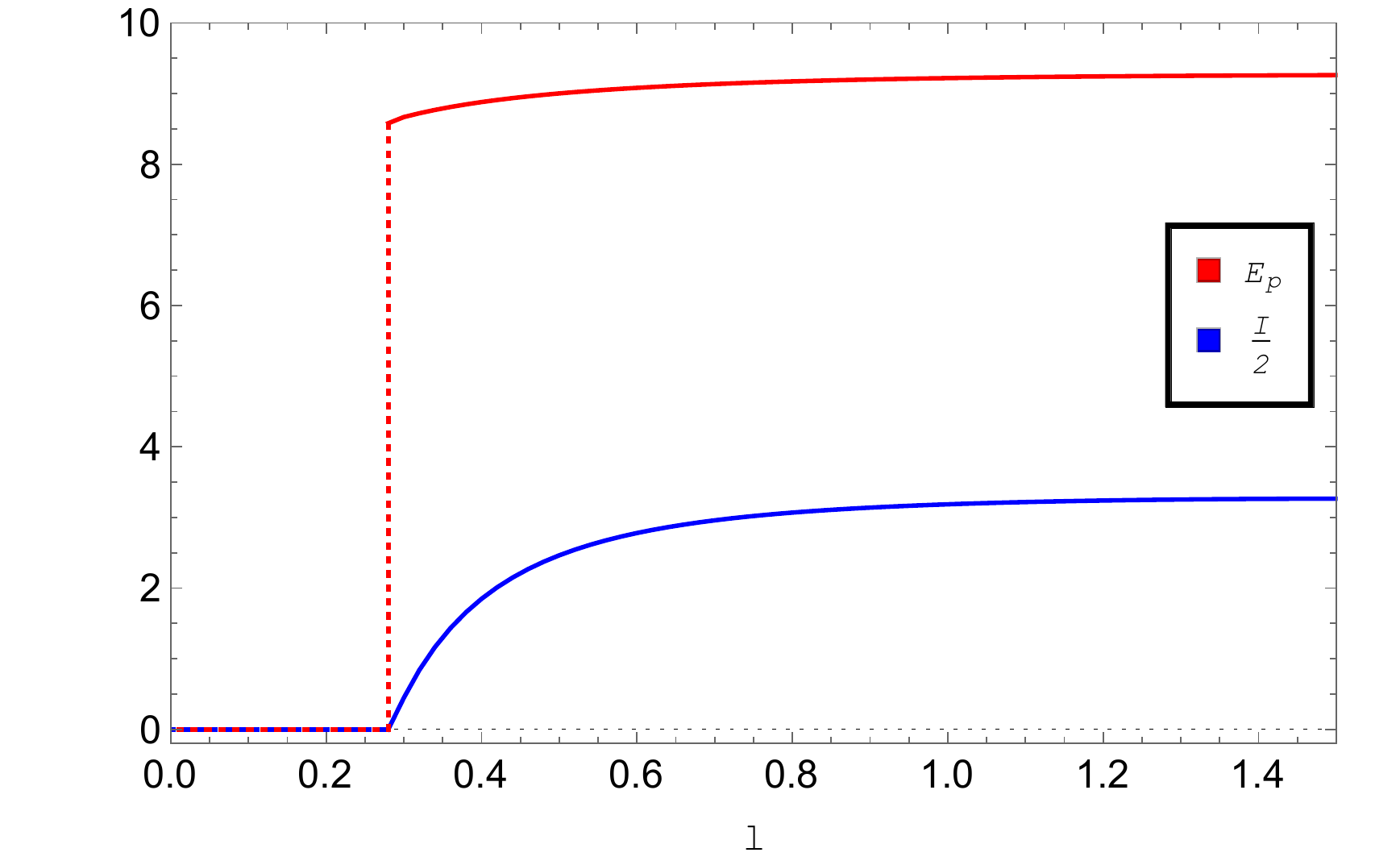} 
\caption{$E_p$ and $I/2$ with respect to $l$ for $l'=0.1$ (left) and $l'=0.2$ (right).}
\label{fig7}
\end{figure} 
\begin{figure}
\includegraphics[width=130 mm]{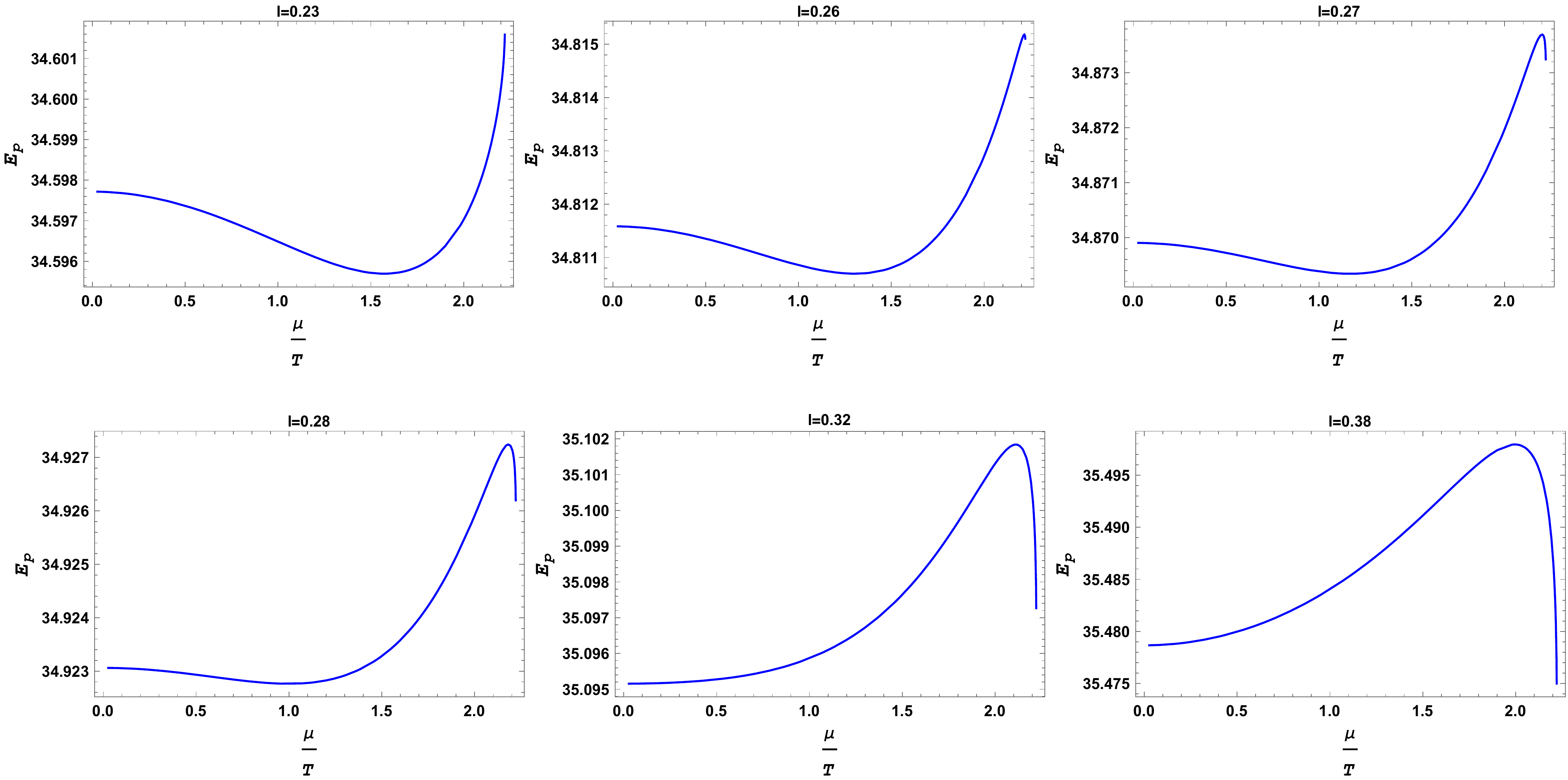}
\includegraphics[width=130 mm]{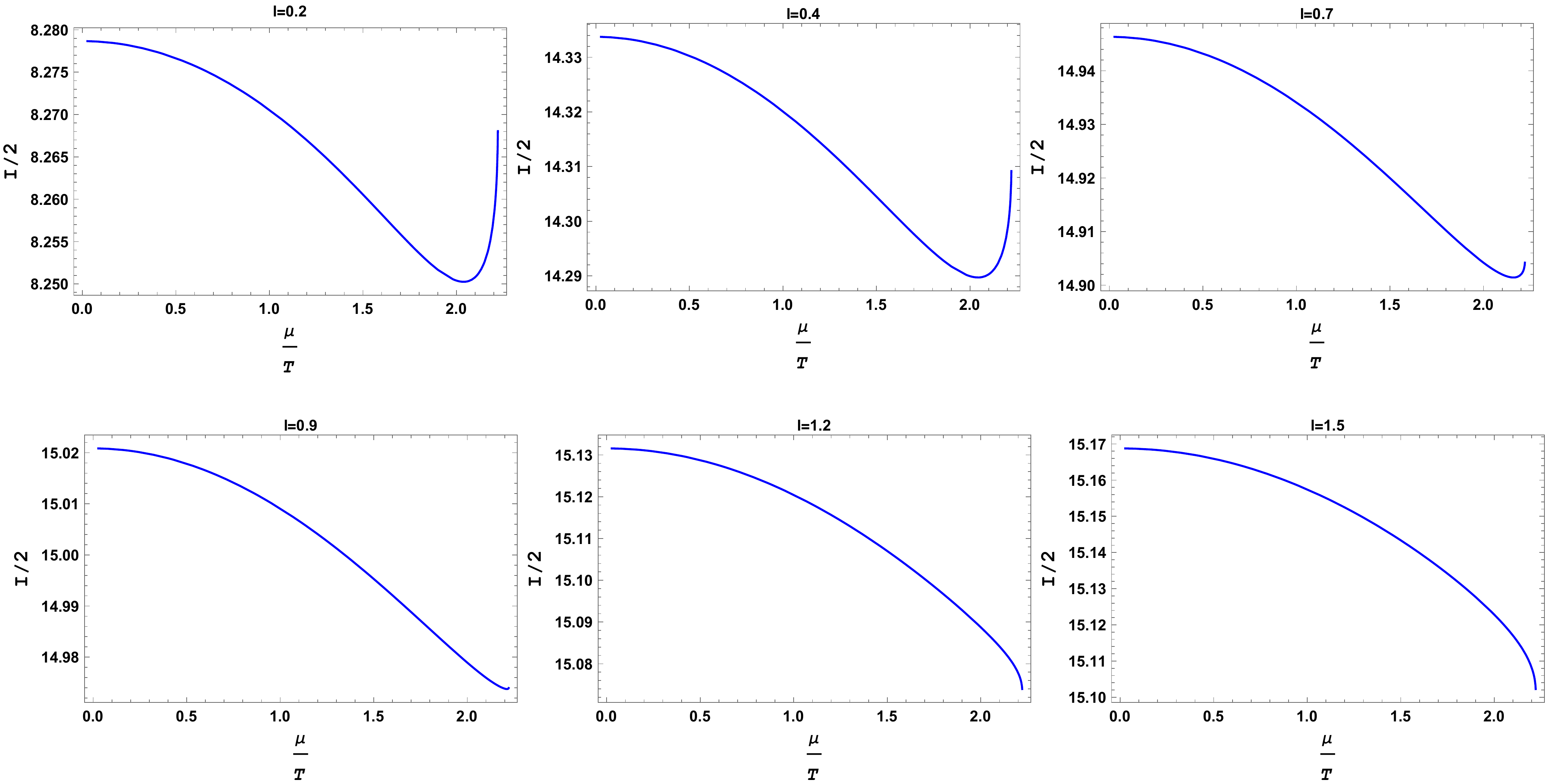} 
\caption{The two top rows: The EoP with respect to $\m$ for $l'=0.1$. The two lower rows: Half of the mutual information, $\frac{I}{2}$, with respect to $\m$ for $l'=0.1$ and different values of $l$. In the all of these diagrams $T$ is fixed and equal to $0.37$. }
\label{fig14}
\end{figure}
\begin{figure}
\centering
\includegraphics[width=75 mm]{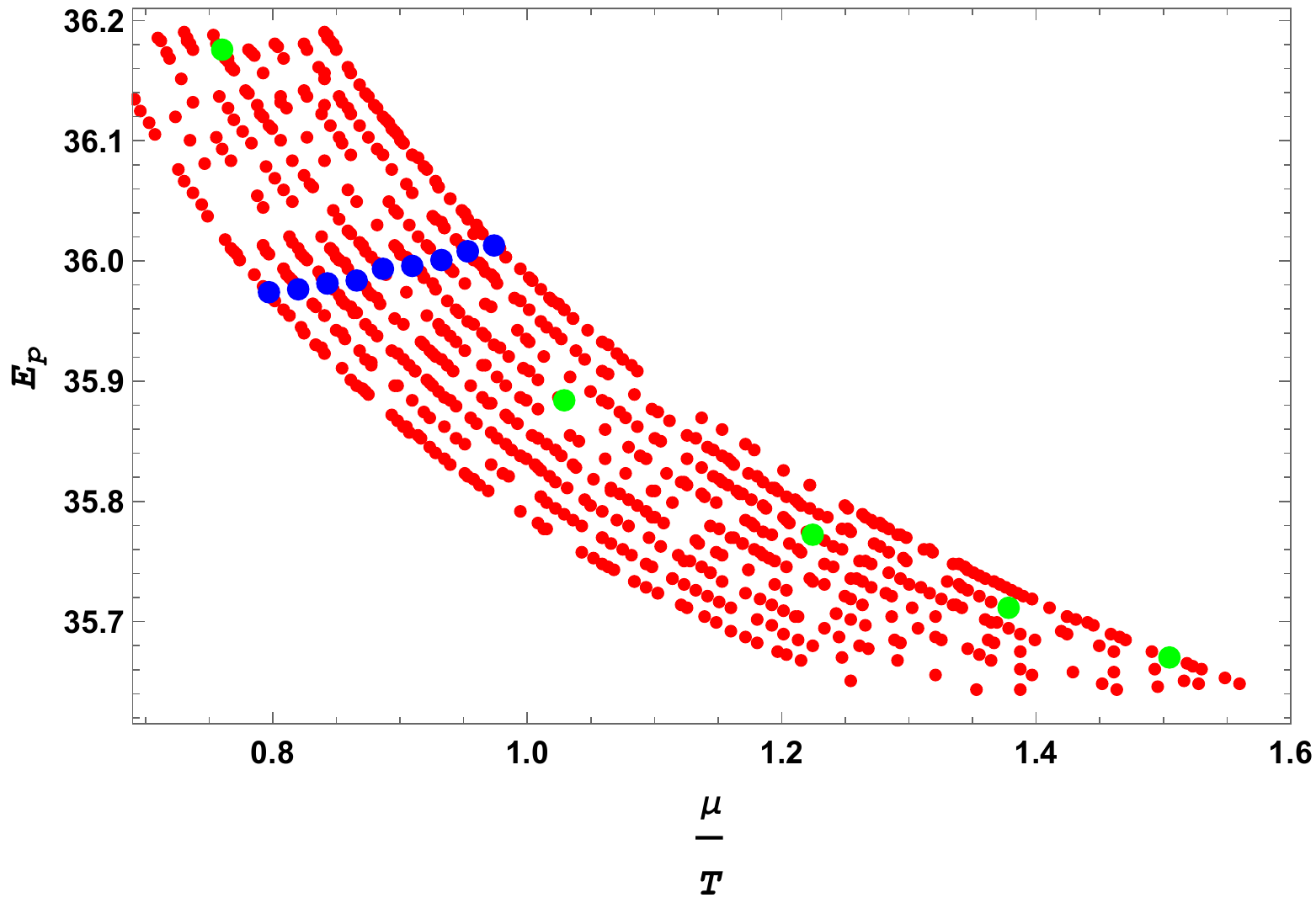}
\includegraphics[width=75 mm]{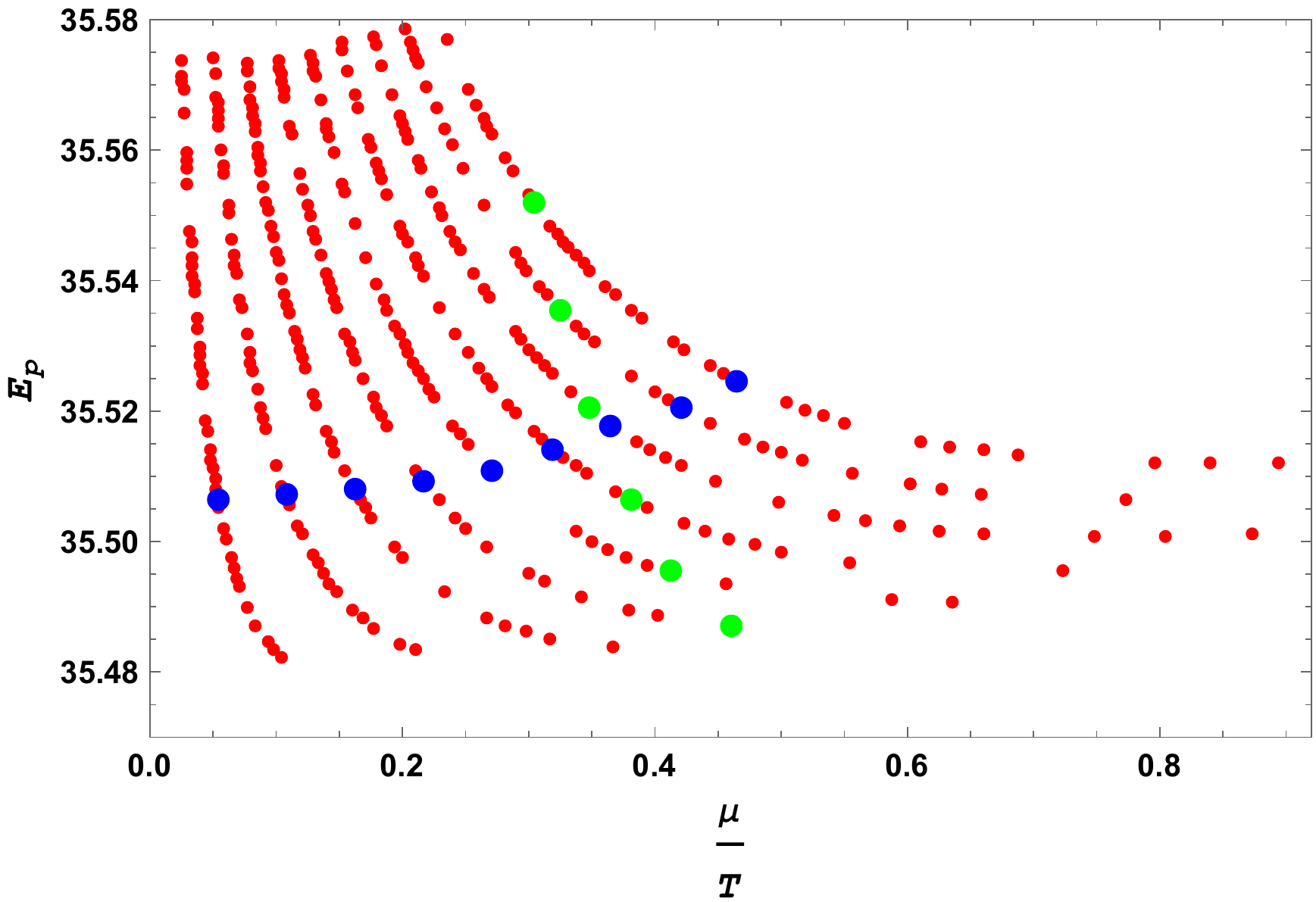}
\caption{Left: The EoP with respect to $\m$ for $l=0.5$ and $l'=0.1$ in the field theory with critical point corresponding to metric \eqref{metric}. Right: The EoP with respect to $\m$ for $l=0.5$ and $l'=0.1$ in the field theory with non-zero chemical potential corresponding to metric \eqref{metric1} with $d=3$. In left (right) figure the green points show the configuration at fixed $\mu=1.077$ ($\mu=0.194$) and blue points show the configuration at fixed $T=1.20$ ($T=0.546$). }
\label{RN2}
\end{figure}  
\section{Numerical Results}
In figures \ref{fig4} and \ref{fig7}, we have checked the inequality between EoP and the mutual information, i.e.
\be\label{ineq}
\frac{I}{2}\le E_p,
\ee
for the two subsystems $A$ and $B$ with the same length $l$. These figures with different values of $l$ and $l'$ convince us that the above relation satisfies for all values of $l$ and $l'$. Clearly, when the mutual information is zero, the EoP is zero as well. Moreover, a phase transition between zero and non-zero values of EoP is observed.

We plot the EoP behavior in terms of $\m$ in figure \ref{fig14}. Although the value of EoP does not substantially depend on the $\m$, for small enough values of $l$ a minimum value of the EoP appears at $\m=(\m)_{min}$. In fact, for $\m>(\m)_{min}$ $(\m<(\m)_{min})$ the value of the EoP increases (decreases) and thus the EoP has a minimum value at $(\m)_{min}$. Moreover, this figure indicates that there exist two values of $\m$ which have the same value of EoP. Near the critical point, the value of EoP grows. Then, by raising $l$ gradually, a maximum value for the EoP appears near the critical point. One should notice that there are values of $l$ for which a maximum and minimum value of EoP exist, simultaneously. For large enough $l$, the minimum disappears and as a result, the EoP has only a maximum at $\m=(\m)_{max}$. In short, we conclude that 
\begin{itemize}
\item The EoP is not a monotonic function of scale $\m$ and as a matter of fact, it experiences distinct behaviors. It can be an increasing or decreasing function of $\m$ depending on the values of $l$ and $l'$ and not $l/l'$.
It is clear that the non-trivial behavior of EoP depends on the values of $l$ and $l'$, though we cannot find their dependence analytically. 
\item The value of $\m$ at which the EoP achieves its maximum or minimum depends on the $l$ and $l'$, as it is expected.
\item In the presence of the chemical potential, the EoP can decrease or increase. In other words, the correlation between two systems becomes stronger or weaker depending on the value of $\m$. 
\item Inequality \eqref{ineq} always satisfies (up to our numerical calculation).
\end{itemize}

An interesting observation of the EoP is that for given values of the subsystems and their separation there are two or three different configurations, labeled by various values of $\m$s, with the same EoP. In fact, for given values of $l$ and $l'$, when the system is described by a mixed state characterized by $\m$ the correlation between the subsystems can be equal independent of $\m$. 
It is then instructive to investigate that this behavior takes place because of the existence of the critical point in the field theory.

In order to check this claim, we consider another charged black hole background corresponding to a field theory in the presence of chemical potential without a critical point. This background has been introduced in \eqref{metric1}. We then plot the EoP in terms of $\m$ in figure \ref{RN2} using the backgrounds metrics \eqref{metric} (left panel) and \eqref{metric1} with $d=3$ (right panel). First of all, it is clearly seen that when $l$ and $l'$ are kept fixed there are many points with different values of $\m$ which have the same value of EoP. From the gauge theory point of view, it means that there exists many mixed states with the same correlation between two subsystems independent of $\m$. It is an important result. Although the states under study are not the same, since they have different values of $\mu$ and $T$, the correlation between two subsystems is identical. Moreover, when $\mu$ ($T$) is kept fixed the EoP decreases (increases) by decreasing temperature (increasing chemical potential). Equivalently, by raising the temperature the correlation increases meaning that two subsystems are strongly correlated with each other at the higher temperature which is reasonable intuitively. At larger $\mu$ the correlation increases too. In short, the EoP or the correlation between the subsystems increases by raising both temperature and/or chemical potential. These results are confirmed for both field theories with and without critical point dual to metrics \eqref{metric} and \eqref{metric1}. Therefore, it seems that the EoP, as a function of $\m$, is not a good observable for distinguishing a critical point between the holographic field theories. 
\begin{figure}
\centering
\includegraphics[width=75 mm]{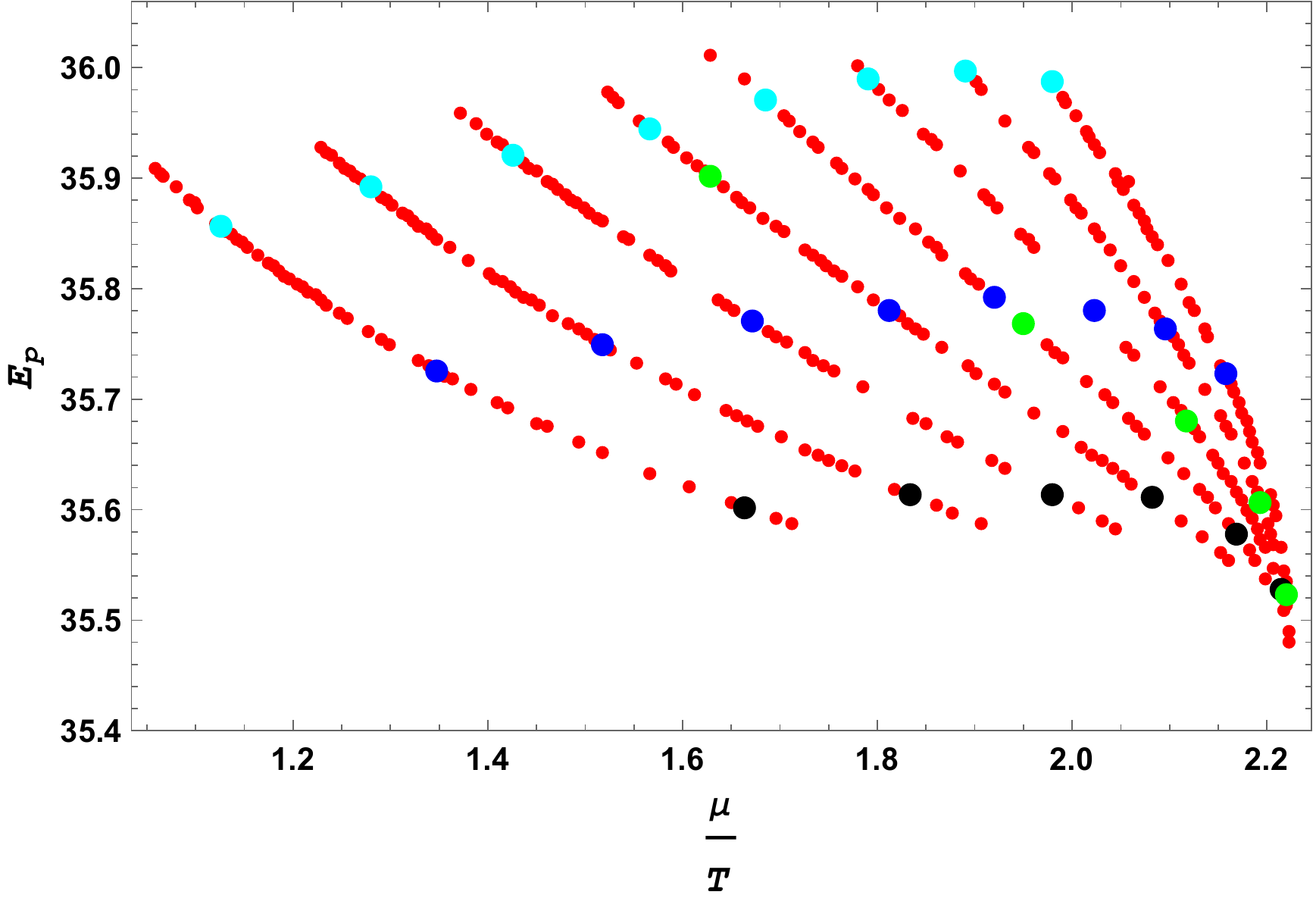}
\caption{The EoP with respect to $\m$ for $l'=0.1$, $l=0.5$ near the critical point. The blue, cyan and black points show $T=0.805$, $0.995$ and $0.61$, respectively. The green points shows $\mu=1.53$.}
\label{RNcritical}
\end{figure} 
\begin{figure}
\centering
\includegraphics[width=75 mm]{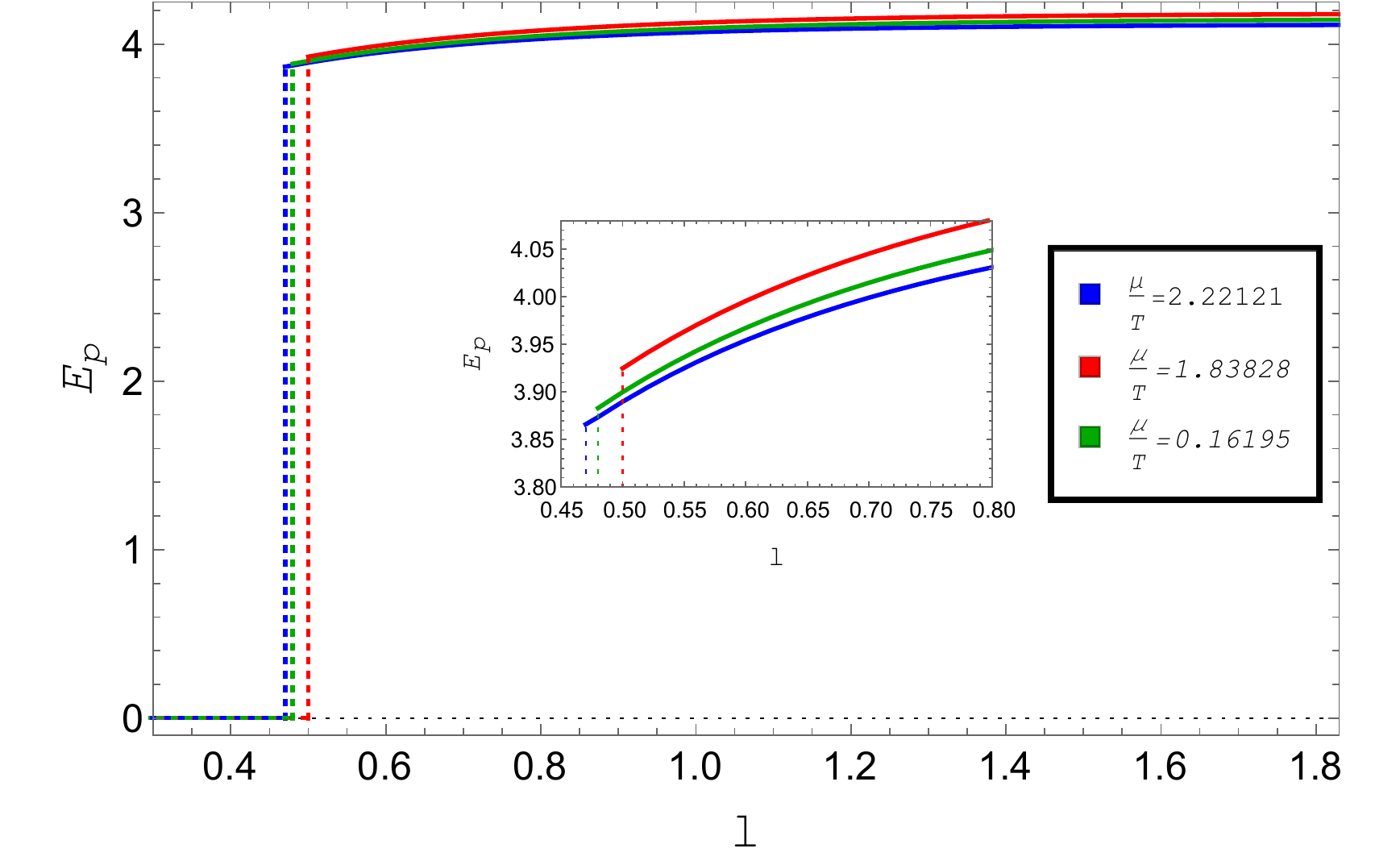}
\includegraphics[width=75 mm]{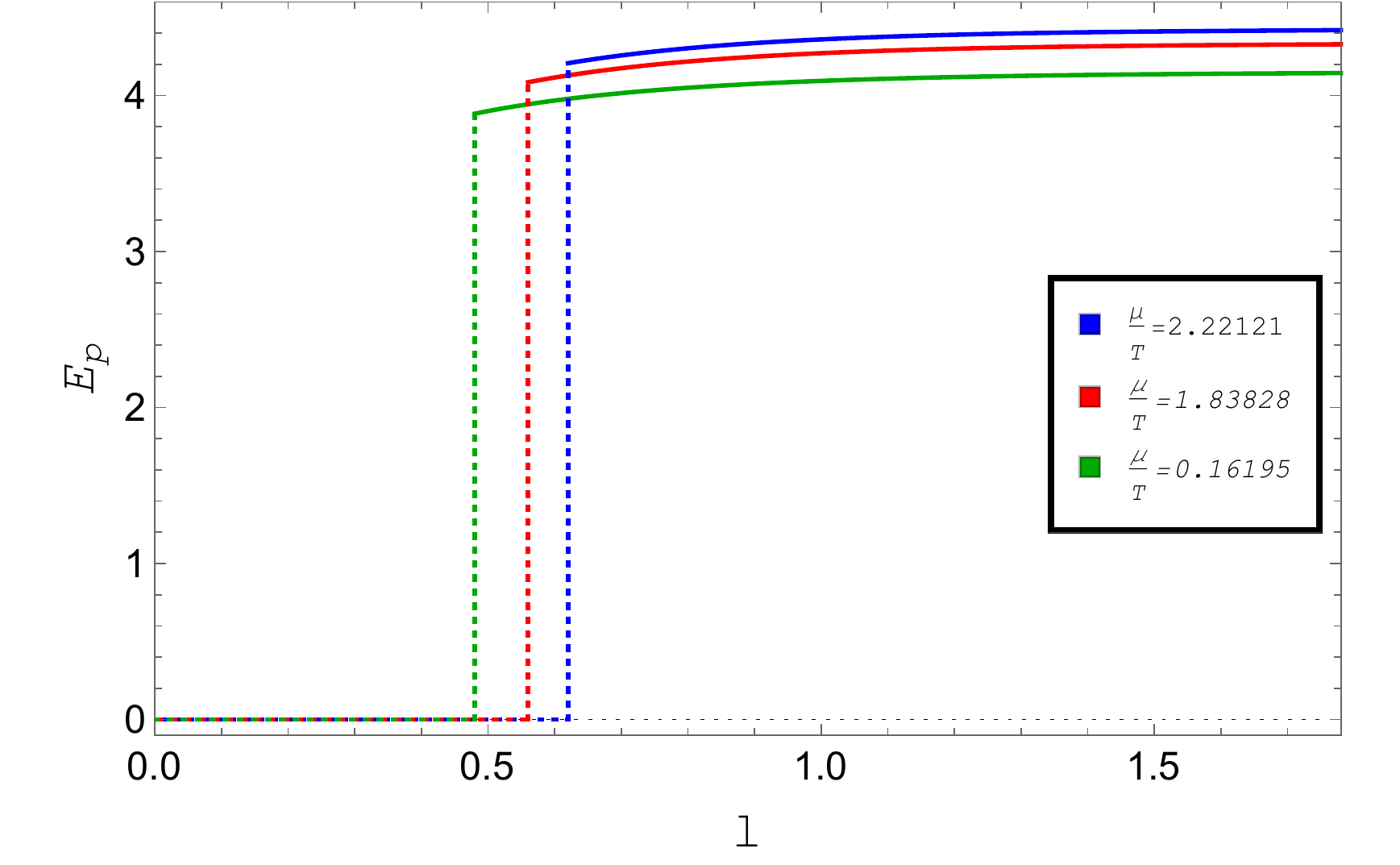}
\caption{Left: The EoP with respect to $l$ for three different values of $\m$ and $l'=0.3$ in the field theory with a critical point. Right: The EoP with respect to $l$ for the same values of $\m$ and $l'$ with non-zero chemical potential corresponding to \eqref{metric1} with $d=3$. In both figures, the well-known phase transition between zero and non-zero mutual information, or equivalently the EoP, is shown by the dashed line. The three different values of $\m$ are chosen to have the same $T$ equal to $0.37$ and different $\mu$.  }
\label{fig19}
\end{figure} 
\begin{figure}
\centering
\includegraphics[width=75 mm]{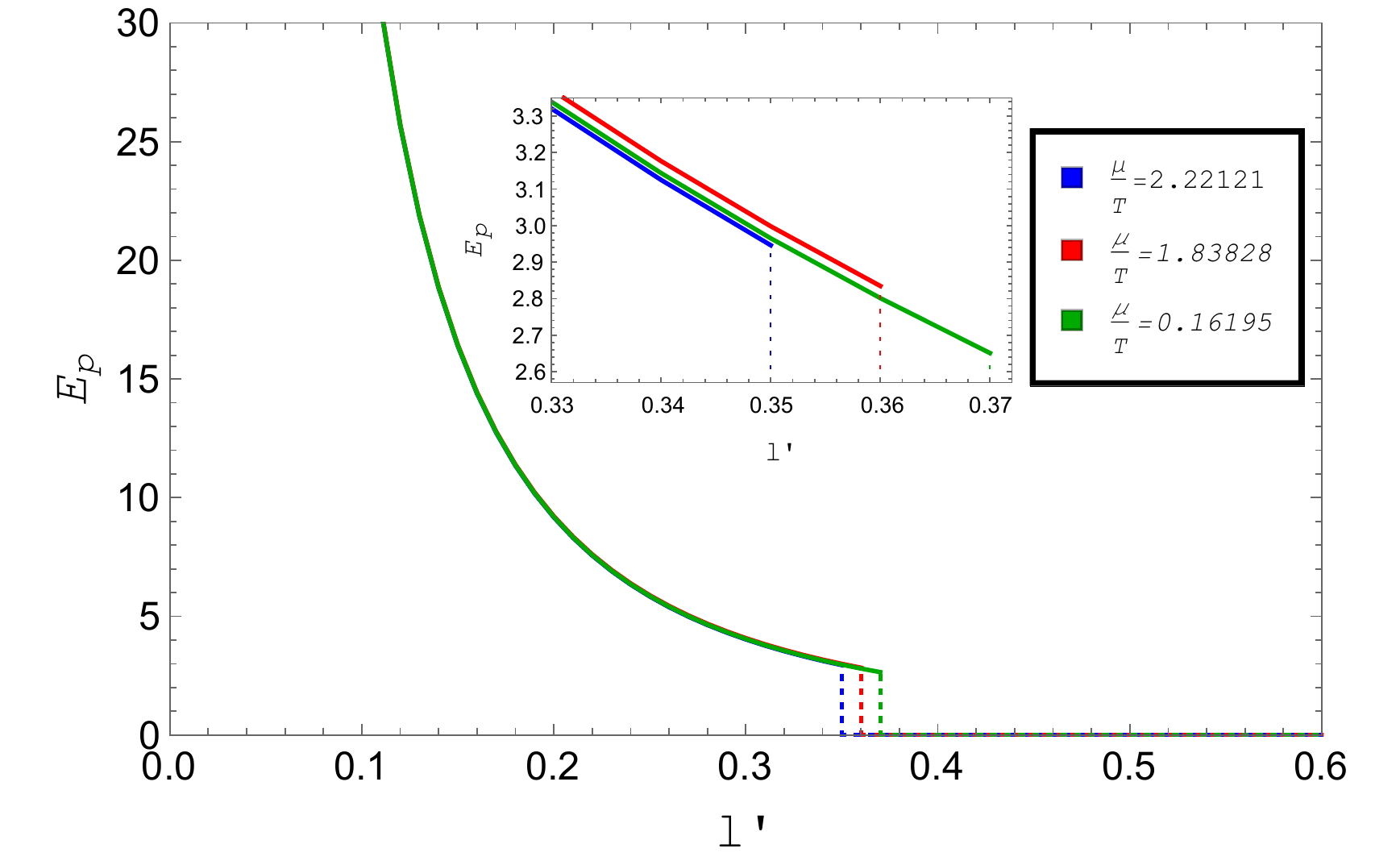}
\includegraphics[width=75 mm]{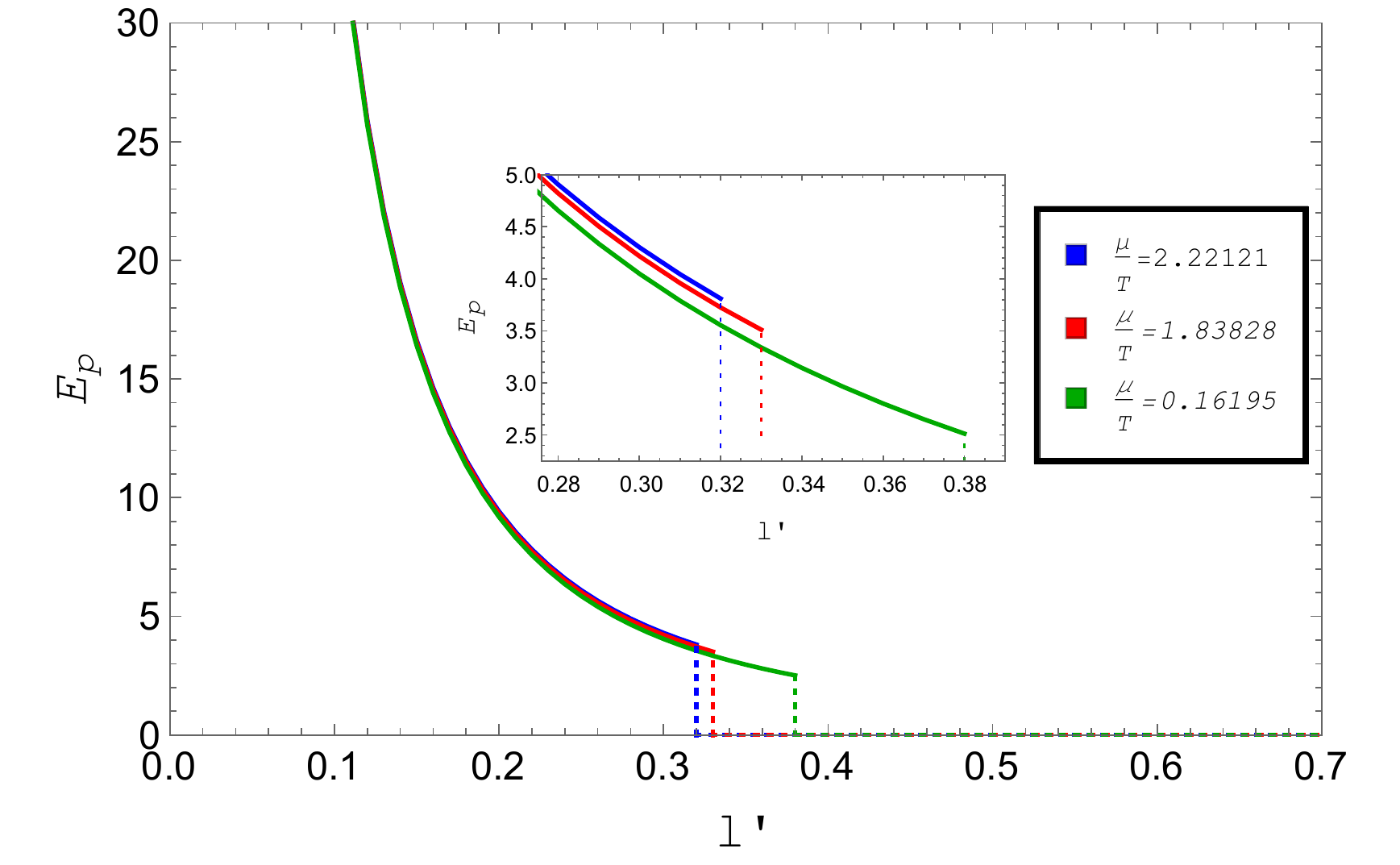}
\caption{The EoP with respect to $l'$ for $l=0.8$. The left (right) panel has been plotted for the field theories dual to \eqref{metric} ((\ref{metric1}) with $d=3$). The three different values of $\m$ are chosen to have the same $T$ equal to $0.37$ and different $\mu$.}
\label{fig22}
\end{figure} 

The EoP in terms of $\m$ near the critical point has been plotted in figure \ref{RNcritical}. This figure indicates that all curves, both fixed temperature and chemical potential curves, converge at $\m=(\m)_*$.  How this behavior is near the critical point is an important question and we will study it later on. As we will see, near the critical point we have $\frac{dE_p}{d(\m)}\propto(\m-(\m)_*)^{-\theta}$ and therefore close to this point the number $\theta$, called critical exponent, describes the variation of the EoP with respect to $\m$. This result is in agreement with the ones reported in \cite{DeWolfe:2011ts, Finazzo:2016psx, Ebrahim:2017gvk, Ebrahim:2018uky, Ebrahim:2020qif}, as we will see.

\begin{figure}
\centering
\includegraphics[width=55 mm]{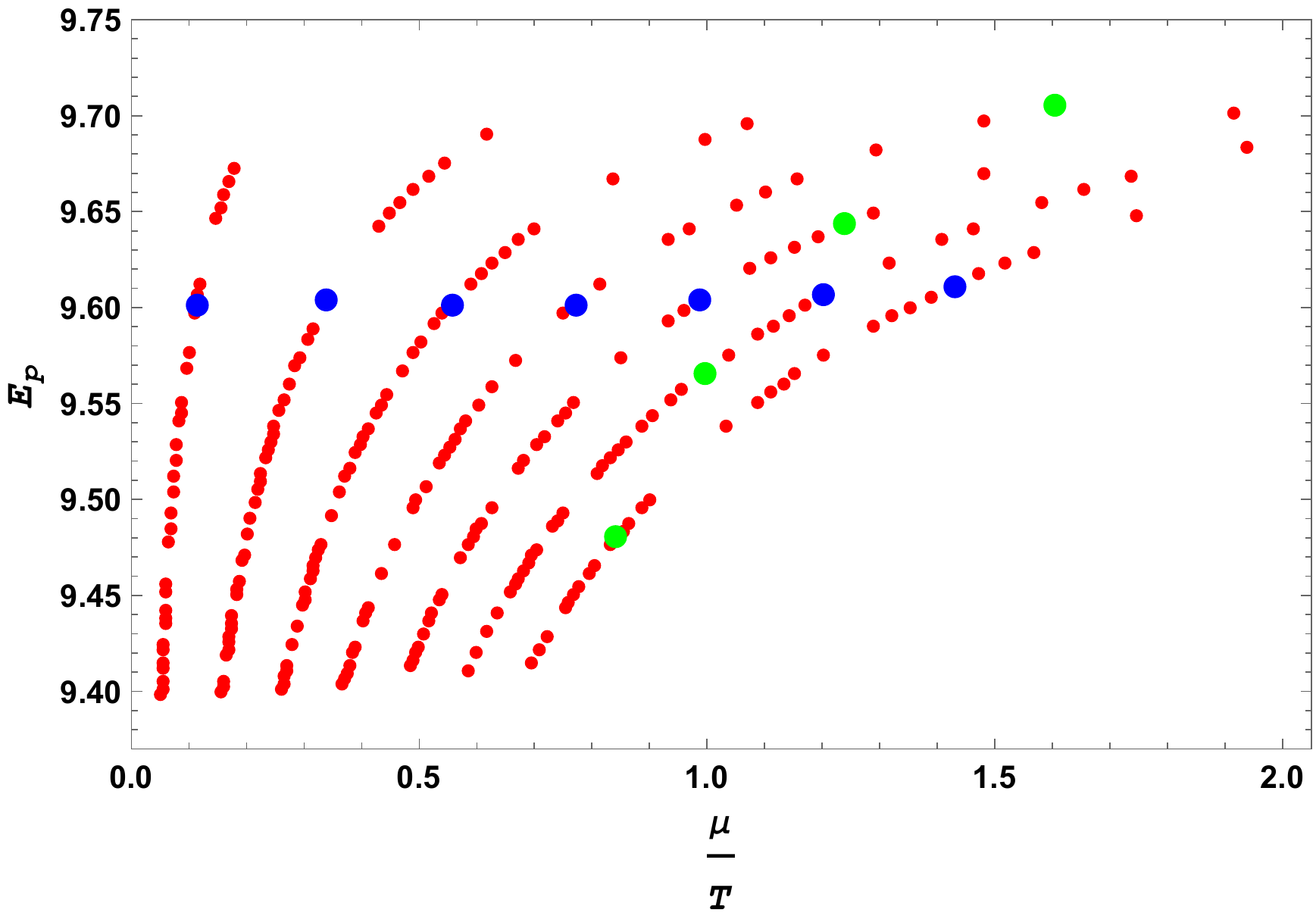}
\includegraphics[width=55 mm]{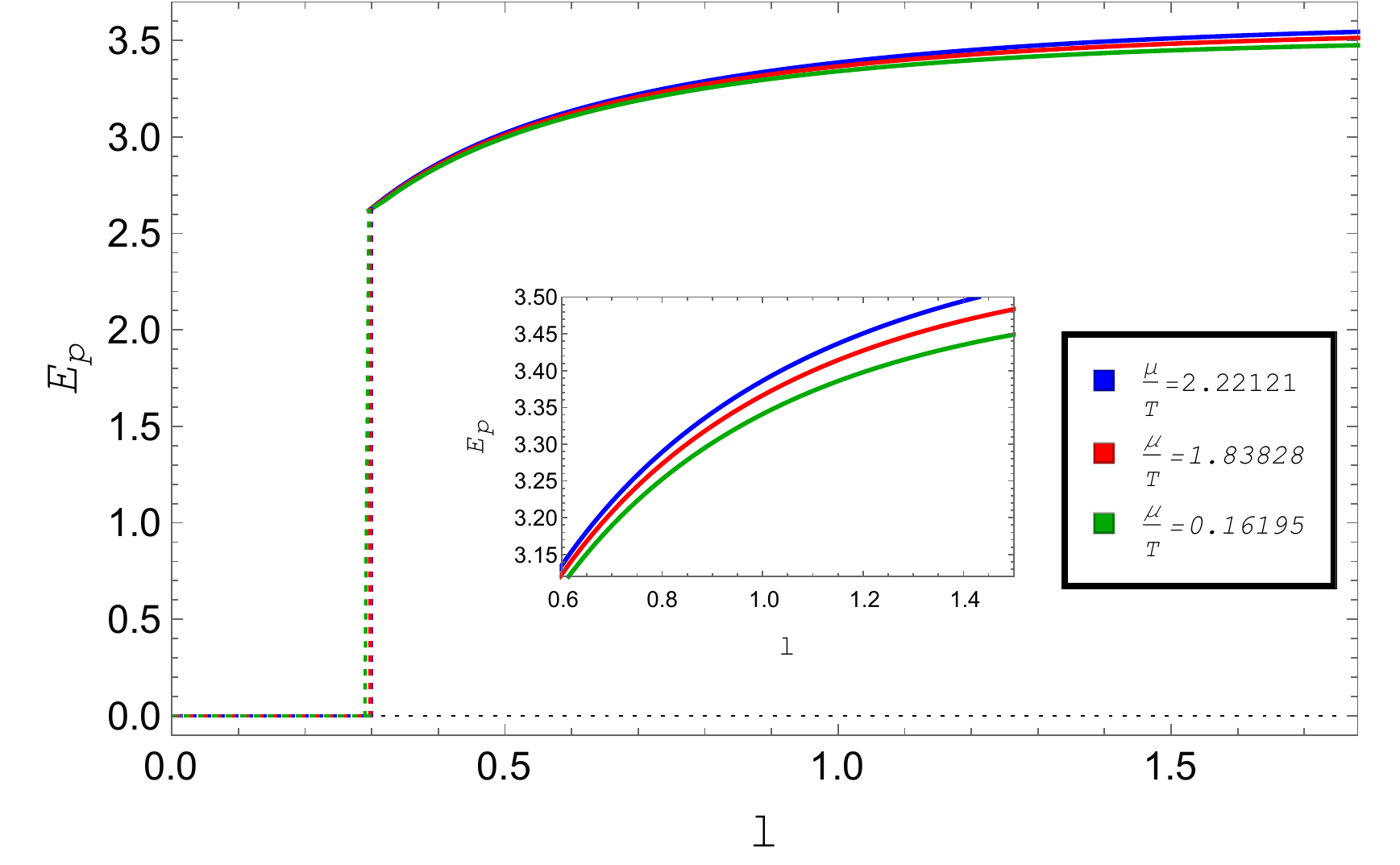}
\includegraphics[width=55 mm]{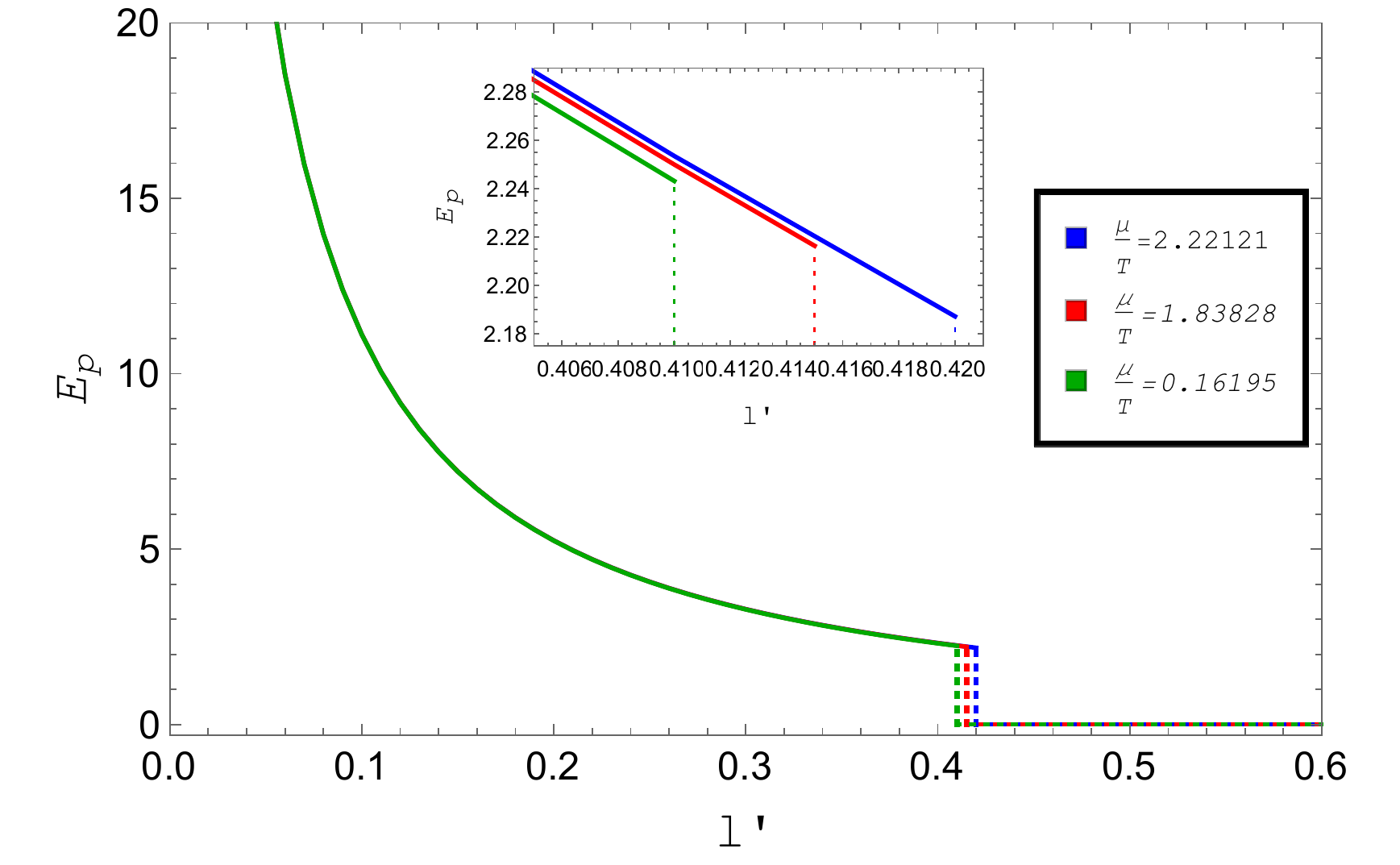}
\caption{Left: The EoP with respect to $\m$ for $l'=0.1$ , $l=0.5$ in 2+1-dimensional field theory. The blue and green points show $T=0.459$ and $\mu=0.508$, respectively. Middle: 
The EoP with respect to $l$ for $l'=0.3$ in 2+1-dimensional field theory. Right: The EoP with respect to $l'$ for $l=0.8$ in 2+1-dimensional field theory. The three different values of $\m$ are chosen to have the same $T$ equal to $0.21$ and different $\mu$.}
\label{RNAdS4}
\end{figure} 

In figure \ref{fig19}, the EoP in terms of distance $l$ has been plotted for three different values of $\m$. Note that $\m=2.22121$ in the left panel specifies a state with $\m$ very close to $(\m)_*$. However, in the right panel, there is no critical point. As it is clearly seen, in the right panel, the EoP and $\m$ increase together. However, in the left panel one can see that the EoP has no general behavior and near the critical point it decreases or increases, see for example the first panel in the first and second row of figure \ref{fig14}. Therefore, the EoP, as a function of $\m$ and $l$, distinguishes which theory has a critical point. Another point we would like to emphasize is that, for large enough $l$, the EoP does not change substantially with distance $l$ for given $\m$ and it is almost constant. It indicates that the degrees of freedom of the field theories at large distance are not strongly correlated with each other.

Figure \ref{fig22} shows the behavior of EoP in terms of $l'$, the distance between two subsystems. The EoP decreases by raising $l'$ and it becomes zero suddenly meaning that the well-known phase transition takes place. In opposition to the previous figure, one can see that although in the left panel there is no general behavior, in the right panel the EoP and $\m$ decrease together. Since the EoP, as a function of $\m$ and $l'$, has different treatments in the field theories with and without the critical point, it can be considered as an appropriate criterion to distinguish between the mentioned theories. It is also significant to notice that the EoP decreases substantially as the distance between subsystems becomes larger. Furthermore, notice that the EoP has the minimum value at the critical point, see figures \ref{fig19} and \ref{fig22}.

We also consider the background \eqref{metric1} with $d=2$. The EoP, \eqref{EoP2} with $d=2$, in terms of $\m$, $l$ and $l'$ has been plotted in figure \ref{RNAdS4}. Our results are similar to the case of $d=3$ and we do not report them here. However, the left panel of above figure shows an opposite treatment compared to the 3+1-dimensional field theories. 
\begin{figure}
\centering
\includegraphics[width=80mm]{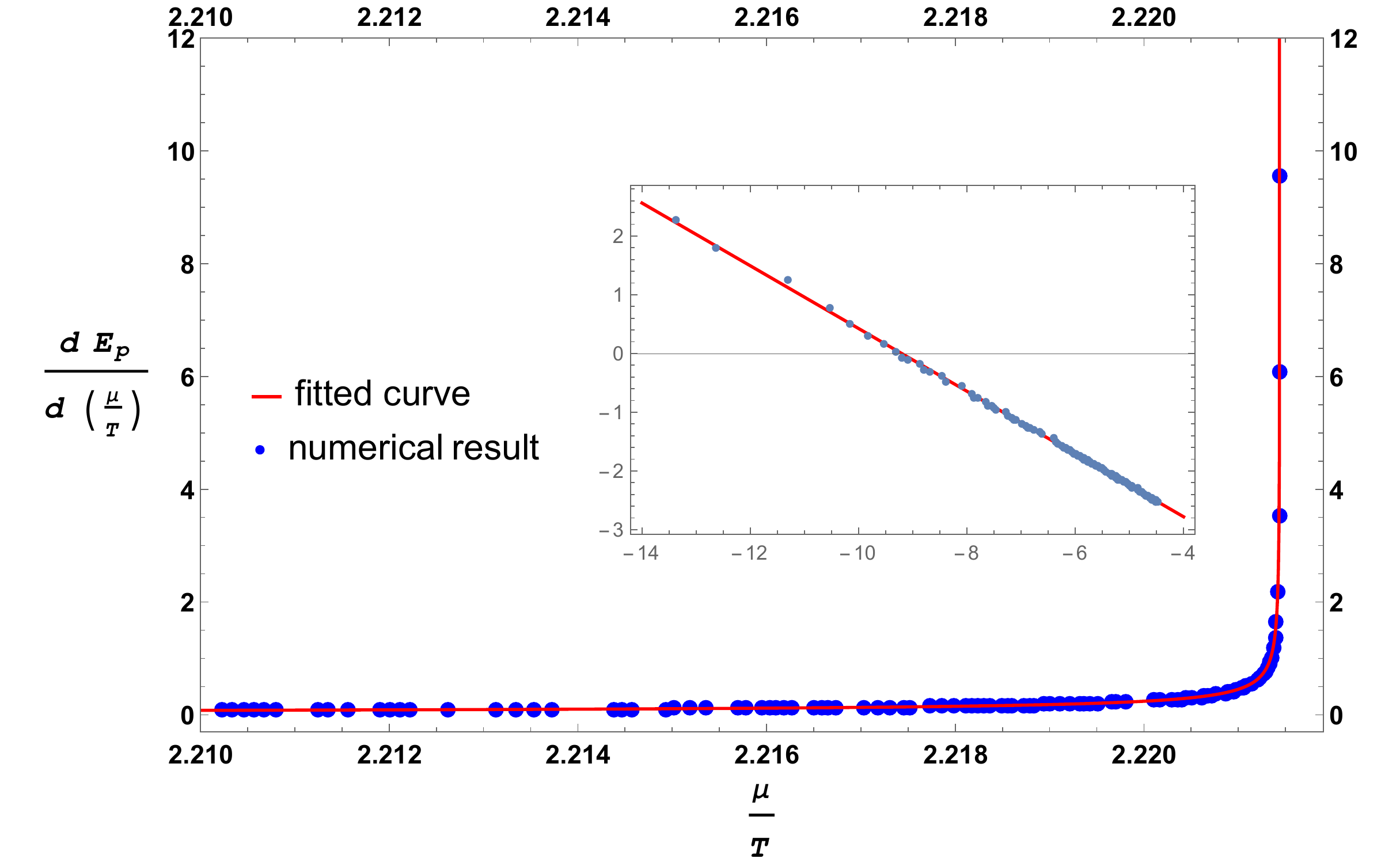}
\includegraphics[width=80 mm]{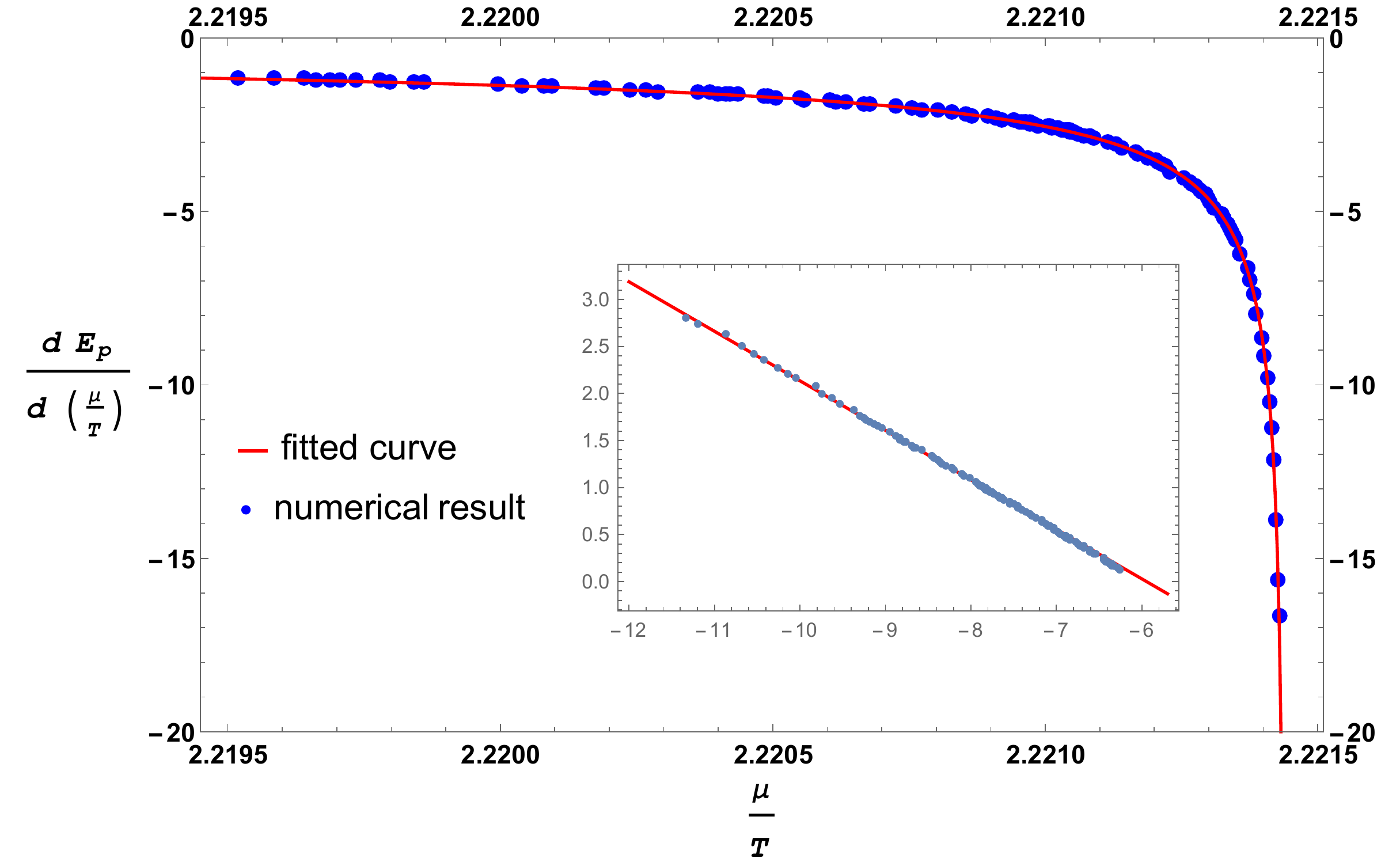}
\caption{The slope of $E_p$ with respect to $\m$. The left diagram has been plot for $l=0.2$ and $l'=0.1$ and right diagram has been plot for $l=0.4$ and $l'=0.2$. For the left (right) figure $\theta$ will be obtained 0.534 (0.526). The corresponding RE and RMS are $0.058$ and $0.068$ for the left panel and $0.053$ and $0.074$ for the right one. Note that the $\frac{dE_p}{(d\m)}$ can be infinitely positive or negative at the critical point. These figures are plotted for $T=0.37$.}
\label{fig30}
\end{figure}
In the last figure, as it was mentioned already, we assume that the slope of EoP with respect to $\m$ changes as $(\m-(\m)_*)^{-\theta}$ where $\theta$ is the critical exponent obtained to be equal to $0.5$ in \cite{DeWolfe:2011ts} using Kubo commutator for conserved currents and confirmed in \cite{Finazzo:2016psx, Ebrahim:2017gvk, Ebrahim:2018uky} by using quasinormal modes, equilibration time and saturation time, respectively. Therefore, we define 
\be
\frac{dE_p}{d(\m)}(i)=\frac{E_p(i+1)-E_p(i)}{\m(i+1)-\m(i)},
\ee
where $i$ refers to the number of numerical data points. Our results indicate that although the quantity we consider here is basically different with above mentioned papers, the value of the critical exponent is around $0.5$, i.e. $0.534$ and $0.526$ for the left and right panel of figure \ref{fig30}. To obtain these two numbers, we also plot the linear log-log diagram for which the critical exponent is the slope of a line, i.e. $\log(\frac{dE_p}{d(\m)})\propto \theta\log(\m-(\m)_*)$. In order to report how well our fitted $\theta$ is we calculate relative error (RE) and root mean square (RMS) which are defined as 
\be\begin{split} 
{\rm{RE}}&=\frac{|\theta-0.5|}{0.5},\cr
{\rm{RMS}}&=\sqrt{\frac{1}{N}\Sigma(y_{fit}-y_{data})^2},
\end{split}\ee
where $y_{fit}$ is the value of fitted function $y$ evaluated at data points $x$ and $y_{data}$ is the corresponding value read from data and $N$ is the number of data points. These numbers are reported in the caption of figure \ref{fig30}.

\end{document}